# Load Balancing with preemptive and non-preemptive task scheduling in Cloud Computing

Mohammad Riyaz Belgaum[1,3], Safeeullah Soomro[1], Zainab Alansari[1,2]

[1]*Department of Computer Studies*
AMA International University, Kingdom of Bahrain.
[2]*Department of Computer Science & Information Technology*
University of Malaya, Kuala Lumpur, Malaysia.
{ bmdriyaz, s.soomro & zeinab }@ amaiu.edu.bh

Shahrulniza Musa[3]
Malaysian Institute of Information Technology
[3]Universiti Kuala Lumpur (UniKL MIIT),
Kuala Lumpur, Malaysia
shahrulniza@unikl.edu.my

Muhammad Alam[4]

[4]*College of Computer Science and IS (IOBM) Karachi, Pak*
Institute of Post Graduate Studies
Universiti Kuala Lumpur (UniKL IPS), KL Malaysia
malam@iobm.edu.pk

Mazliham Mohd Su'ud[5]
[5]Malaysia France Institute
Universiti Kuala Lumpur, (UniKL MFI),
Bandar Baru Bangi, Malaysia
mazliham@unikl.edu.my

*Abstract* – In Cloud computing environment the resources are managed dynamically based on the need and demand for resources for a particular task. With a lot of challenges to be addressed our concern is Load balancing where load balancing is done for optimal usage of resources and reduces the cost associated with it as we use pay-as-you-go policy. The task scheduling is done by the cloud service provider using preemption and non-preemption based on the requirements in a virtualized scenario which has been focused here. In this paper, various task scheduling algorithms are studied to present the dynamic allocation of resources under each category and the ways each of this scheduling algorithm adapts to handle the load and have high-performance computing.

*Index Terms – Load Balancing, Scheduling, Virtualization, Dynamic Allocation.*

## I. Introduction

Cloud Computing is an environment where the computing resources are used from the available pool, by the clients on demand [1]. In other words, all the needed resources are tapped to use by the clients instead of managing them. The significant advantage of pay-as-you-go enables the enterprises to reduce the sunk costs of maintaining the physical resources which are left unused. The resources are allocated dynamically [2] to the tasks based on the scheduling algorithm used. As there are different types of environments like preemptive and non-preemptive, the load balancer takes into consideration the various parameters like the need, availability, scheduling policy, user requirements, bandwidth, frequency, QoS, etc.

Dynamic load balancing techniques are used to have fault tolerant systems [3], as the quantity of tasks increase and the virtual machines are limited in a real environment. The overloaded virtual machine causes failure of the task completion making the utilization of resource go in vain. The virtualization characteristic of the cloud computing enables the user to be free from the underlying work done by the other components in the cloud computing. Various virtual machines are grouped into clusters. Virtual machines that perform similar tasks or that are assigned based on homogeneity can be a grouped into clusters. In terms of security, the most secure cloud service provider is the one in which the presence of physical server is not known to the virtual instance that is running. If this is the case, then a determined hacker faces a tough time which is targeting to hack the particular task present in the physical environment. The task needs to have a Service Level Agreement [4] with the third parties to be more secure. The information when is to be accessed by a wide variety of users by using IoT [5] in different domains like in health care etc., it should be readily available.

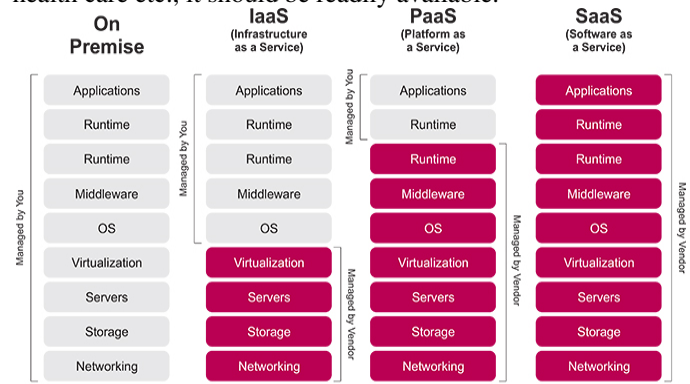

Fig. 1 Cloud Computing Service Models.

The above figure shows the various cloud computing service models and the one responsible for managing them. The ability to understand and to fully trust the availability, reliability, and performance of the cloud are the key reasons for the organizations to show their interest to move to the cloud. For all the service models, still, the aim of each service model is to serve the client in an interrupted way by providing the best services in an efficient manner. Load balancing is the primary concern that needs the load to be balanced by dynamically allocating the resources to each required node and ensures to maintain the QoS.

## II. Related Literature

Various Load Balancing algorithms have been studied in detail, like one of them being implemented on checkpoint based [6]. The cloud services were ranked based on the checkpoint based load balancing considering the user's requirements and maintaining the QoS.

The authors in [7] have discussed how cloud computing applications have extended their services in combination with a rapid and fast moving communication media know as mobile cloud computing. The architectures, characteristics, similarities, and challenges related to both of the domain have been discussed there. Further, in [8], the authors have shown how HTML5 is used to implement the applications and services of cloud in an efficient manner. Still, the gaps between traditional cloud computing and mobile cloud computing were shown.

One of the preemptive task scheduling being priority scheduling was used in [9] to maximize throughput, where they calculated the energy consumption for scheduling the jobs on computing servers. Considering the requirements and server frequency, one of the available computing servers was allocated following the best fit policy and using DVFS. Job scheduling in the Green cloud is categorized on load balancing, temperature, and energy efficiency.

The authors in [10] have presented a scheduling strategy based on genetic algorithm for the load balancing of resources in cloud computing. The results showed that it achieved the best solution with least or no migration. While in real cloud computing where there are dynamic changes in the virtual machines and the computing cost increases with the unpredicted load, it was proposed to have a mechanism to monitor and analyze the problem of load balancing.

In [11], the authors have proposed a load balancing algorithm based on honey bee behavior. This algorithm was tested on the independent tasks which are non-preemptive. On comparison with other existing algorithms, the results showed an improvement in the execution time and also reducing the waiting time in the queue while considering the QoS.

In order to address the online scheduling problem in IaaS, a preemptive scheduling algorithm with task migration is proposed in [12]. The higher priority task is made to wait for a long time and sometimes misses the deadline, affecting the system performance and response time. The proposed algorithm showed an improvement on the performance-maximizing the utility of resources over traditional scheduling.

The tasks in real time scheduling are scheduled non-preemptively to utilize the resources fully. The challenges related to online video streaming and cloud-based gaming in a shared virtualized environment were studied in [13]. These were used in the resource management as the benefitable task was selected for execution and a non-benefitable task was delayed in execution.

From the various challenges available right in front in cloud computing, one of the challenges of load balancing was considered in [14], where the jobs arrive according to stochastic process and the virtual machines are to be assigned focusing on the resource allocation problems. In the non-preemptive environment, the virtual machines used frame based processes. Simulative results show that the long frame duration processes are good to have a better throughput.

A pricing in combination with resource allocation was used in [15] by the authors. The cloud service provider allocates the jobs considering the characteristic of the flexibility of cloud computing to the resources needed. A resource allocation was proposed in a non-preemptive environment to avoid checkpoints as the jobs which could complete execution in one shot were selected.

In [16], the authors used Haizea, as a resource manager to dynamically allocate the resources to the leases considering the four policies namely immediate, best effort, advanced reservation and deadline e sensitive. As the resources were to be allocated dynamically in a preemptive computing, swapping and backfilling was added to reschedule the processes.

The internal data centers were preferred first in migrating the task to external data centers in [17]. Though linear programming can be used to tackle such problems, a mathematical analysis using binary integer program formulation was proposed for scheduling and cost was evaluated considering the key parameters in them. The results showed that it was suitable for the public cloud model but was not appropriate for the hybrid cloud model.

The bandwidth requirements were considered in [18] for task scheduling in cloud computing, and the authors focused on divisible load applications by processing the tasks in parallel independently. A heuristic algorithm named Bandwidth-aware task scheduling (BATS) was proposed. It was simulated and compared with the other task scheduling algorithms, and the results proved that it could achieve better performance over others.

The dynamic priority scheduler [19] used was more fruitful than the Hadoop schedulers in performing on the number of queues. Except for the limitation of memory capacity and having higher level SLA, it ruled out the performance of fair share scheduling. This mechanism helped the scheduler in prioritizing the users and the tasks for processing.

A modification to the Max-Min algorithm was proposed in [20] for scheduling the tasks. Its implementation was compared by the Min-Min, Max-Min, and RASA algorithms. A lower makespan was observed in comparison with the other algorithms by improving the existing Max-Min algorithm in cloud computing.

A selective algorithm was proposed for resource provisioning to the users based on the requirements in [21]. The goal was to minimize the overall makespan of the tasks on the virtual machines and provide a better QoS. Min-Min and Max-Min algorithms were studied on certain common criteria like the resource capabilities, cloudlet file size was implemented in space shared and time shared modes. The proposed algorithm performed better than the other algorithms considered here in comparison by improving the throughput and minimizing makespan.

III. DISCUSSION

The availability of the resource describes how often it can be used over a period. For example, if a resource is accessible to the accomplish a particular task for 59 minutes out of 60 minutes, then it is said that the availability rating is 98.33% for that resource. So it is always expected for a system to have high availability of resources.

A. *Estimating the availability:*

The availability of a resource can be mathematically calculated as:

$$a^e = (mp - (r^l * d^e))/mp \quad (1)$$

Where, $a^e$ is the expected availability of the resource, $mp$ is the measurement period, $r^l$ is the likelihood of resource loss in a given period, $d^e$ is the expected downtime from loss of the resource.

The resources in a cloud computing environment can be CPUs, memory storage, networks, services, and applications. The scheduling policies manage the resources. There are various scheduling policies categorized as preemptive and non-preemptive scheduling. The resources needed to complete the task are estimated, and they are checked for availability using the above formula. An efficient scheduling algorithm is the one who finds the resources always available with no other task waiting in the queue for its completion. Cloud Analyst, is a simulator tool used here to analyze the results. Some of the scheduling policies like Round Robin(RR), Equally spread current execution load(ESCE), throttled are used for simulation. Various loads are distributed on the data centers in different regions, with user base assigned to each data center. The metrics used are response time, data center servicing times and data center loading.

| Userbase | Avg (ms) | Min (ms) | Max (ms) |
|---|---|---|---|
| UB1 | 50.208 | 37.606 | 60.858 |
| UB2 | 49.958 | 38.641 | 60.398 |
| UB3 | 49.921 | 38.881 | 61.058 |
| UB4 | 50.211 | 39.136 | 59.639 |
| UB5 | 50.205 | 37.657 | 60.658 |
| UB6 | 50.154 | 38.881 | 60.632 |

Fig. 2 Response time by region using RR policy.

From fig 2, it is clear that for each userbase a data center is allocated and so the average response times are almost close to each other.

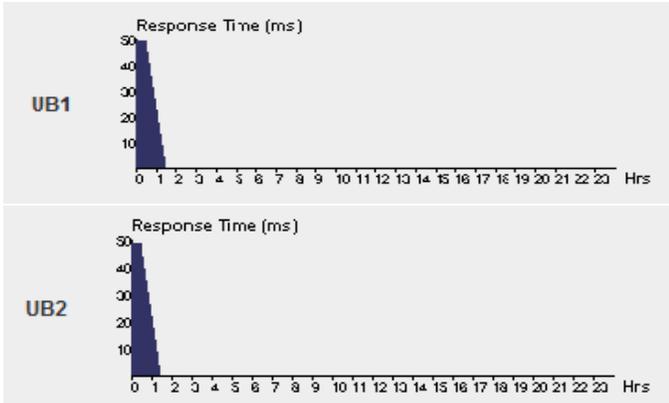

Fig. 3 Response time at each userbase using RR policy.

| Data Center | Avg (ms) | Min (ms) | Max (ms) |
|---|---|---|---|
| DC1 | 0.463 | 0.016 | 0.856 |
| DC2 | 0.502 | 0.028 | 0.891 |
| DC3 | 0.486 | 0.016 | 0.88 |
| DC4 | 0.501 | 0.017 | 0.883 |
| DC5 | 0.497 | 0.028 | 0.896 |
| DC6 | 0.493 | 0.017 | 0.881 |

Fig. 4 Service times at each data center using RR policy.

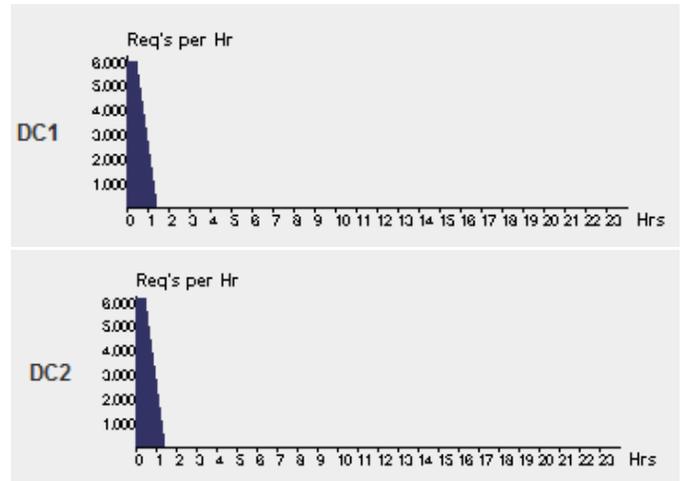

Fig. 5 Data center loading at each data center using RR policy.

Using the Round Robin policy, the response time at each user base can be seen in fig.3, the processing times at each data center can be seen in fig.4 and the loading on each of the data center per hour can be seen in fig.5. Because the policy used is Round Robin which is preemptive, gives a fair share to each task without making any task to wait for a long time in the queue. However, the problem with this policy is that even a small task has to wait for its turn if it is the last one to be processed.

| Userbase | Avg (ms) | Min (ms) | Max (ms) |
|---|---|---|---|
| UB1 | 50.18 | 40.106 | 60.609 |
| UB2 | 50.161 | 39.145 | 61.645 |
| UB3 | 50.13 | 38.63 | 60.129 |
| UB4 | 50.031 | 37.624 | 60.848 |
| UB5 | 50.163 | 39.912 | 62.149 |
| UB6 | 49.834 | 39.882 | 60.382 |

Fig. 6 Response time by region using ESCE policy.

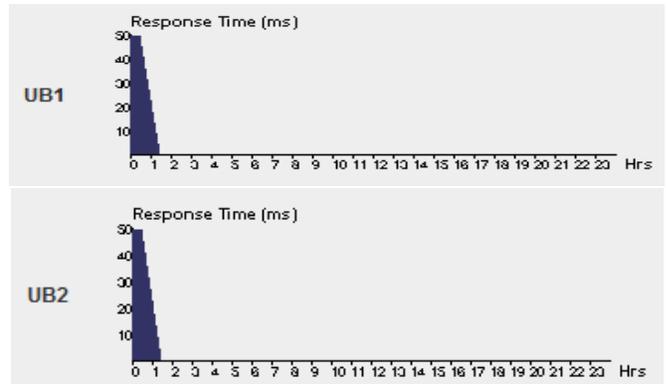

Fig. 7 Response time at each userbase using ESCE policy.

| Data Center | Avg (ms) | Min (ms) | Max (ms) |
|---|---|---|---|
| DC1 | 0.465 | 0.014 | 0.856 |
| DC2 | 0.491 | 0.014 | 0.89 |
| DC3 | 0.488 | 0.017 | 0.879 |
| DC4 | 0.497 | 0.031 | 0.883 |
| DC5 | 0.485 | 0.026 | 0.9 |
| DC6 | 0.493 | 0.03 | 0.881 |

Fig. 8 Service times at each data center using ECSE policy.

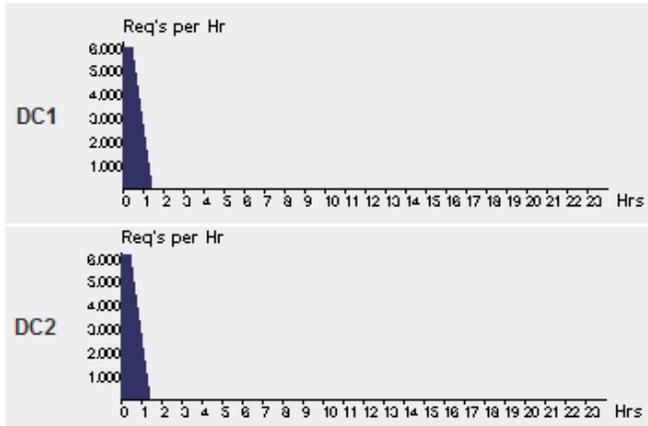

Fig. 9 Data center loading at each data center using ESCE policy.

With equally spread current execution load policy, the user bases are assigned to each data center equally with no data center overloaded with tasks at the very beginning. So there is no preemption here. As there is no preemption, the results in fig. 6, show that the average response time at each user base in each region is less than the response time in other policies. Fig.7 graphically represents the response time in milliseconds. Fig.8 shows the processing time is also less than the service times in other policies. Data center loading is seen in Fig.9 serving the number of requests per hour.

| Userbase | Avg (ms) | Min (ms) | Max (ms) |
|---|---|---|---|
| UB1 | 50.485 | 37.652 | 60.932 |
| UB2 | 50.669 | 39.791 | 61.796 |
| UB3 | 50.651 | 39.881 | 61.298 |
| UB4 | 50.738 | 39.186 | 60.888 |
| UB5 | 50.776 | 38.807 | 61.706 |
| UB6 | 50.793 | 40.256 | 62.031 |

Fig. 10 Response time by region using throttled policy.

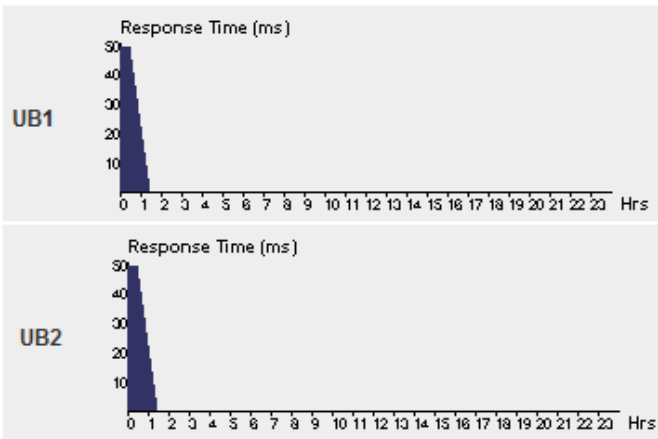

Fig. 11 Response time at each userbase using throttled policy.

| Data Center | Avg (ms) | Min (ms) | Max (ms) |
|---|---|---|---|
| DC1 | 0.762 | 0.024 | 1.89 |
| DC2 | 1.183 | 0.058 | 7.51 |
| DC3 | 1.183 | 0.031 | 5.603 |
| DC4 | 1.04 | 0.032 | 2.178 |
| DC5 | 1.08 | 0.032 | 3.712 |
| DC6 | 1.15 | 0.044 | 3.452 |

Fig. 12 Service times at each data center using throttled policy.

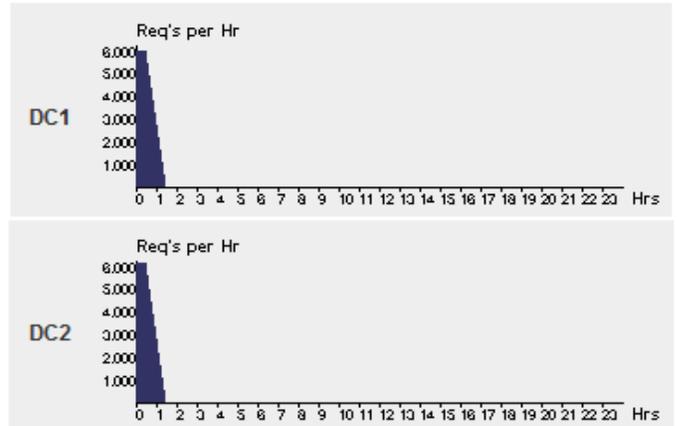

Fig. 13 Data center loading at each data center using throttled policy.

In throttled scheduling, some of the tasks are killed at the data center, and they are migrated to other for processing without the knowledge of the user. The results in fig. 10 show the response times at each user base by region, and it is observed that the response time is more here as some of the tasks are migrated to other data center. Fig.11 shows the response time at each user base graphically. The processing time taken by each data center can be observed in fig.12, where it is more than the other policies. In fig.13 the data center loading is shown which is also more as the tasks are throttled based on the load at each data center and migrated to other data centers for execution.

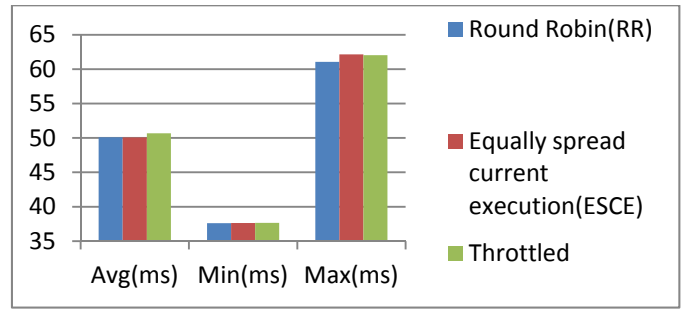

Fig. 14 Response times of all policies.

The response times of all the policies are graphically represented in fig. 14, which shows that equally spread current execution policy has an average less response time but the max response time is more than others.

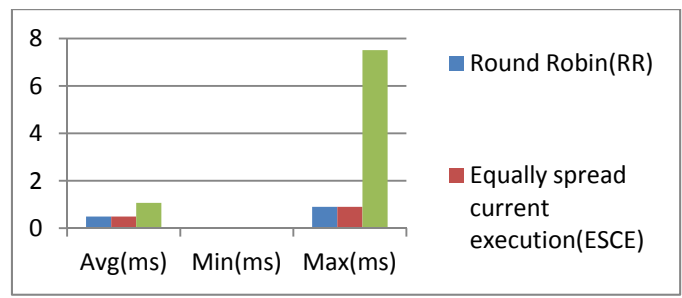

Fig. 15 Data Center processing times of all policies

The data center processing time is more in throttled policy when compared with other scheduling policies as shown in fig. 15. With all these scheduling policies the virtual machines are allocated to the tasks, and still, more such policies can be added to optimize the response time and to lower the overload on the data centers.

## IV. CONCLUSION

This study is conducted to schedule the tasks in the cloud computing environment using preemptive and non-preemptive policies. The availability of the resource is first known, and then it is allocated. Load balancing is a very challenging task to be addressed in cloud computing reduces the overhead on the part of data centers and can utilize that time in processing. The response time at each user base, processing time at each data center using Round Robin Equally spread current execution, throttled policies are compared. The simulative results show that the metrics used help us to conclude that an equally divided current execution load policy is a better one for the data centers to use in located in different regions.

## V. Future Studies

Undoubtedly, this research can be the basis for making other scheduling policies to reduce the response time, processing time and reduce the load on the data centers. Further, it can be extended to other policies and other metrics can be used to have a more optimized policy.